\documentclass[12pt]{iopart}
\usepackage{iopams}  
\begin{document}

\title[Magnetic properties of HfO$_2$ thin films]{Magnetic properties of HfO$_2$ thin films.}

\author{N~Hadacek\dag,~A~Nosov\ddag,~L~Ranno\dag,~P~Strobel\dag~and~R-M~Gal\'{e}ra\dag~}

\address{\dag\ Institut N\'{e}el, CNRS-UJF, BP 166, 38042 Grenoble Cedex 9, France}

\address{\ddag\ Electric Phenomena Lab., Institute of Metal Physics, Ural Division of the Russian Academy of Sciences, S. Kovalevskoi str.18, GSP-170, 620041, Ekaterinburg, Russian Federation}
\ead{rose-marie.galera@grenoble.cnrs.fr}

\begin{abstract}
We report on the magnetic and transport studies of hafnium oxide thin films grown by pulsed-laser deposition on sapphire substrates under different oxygen pressures, ranging from 10$^{-7}$ to 10$^{-1}~mbar$. Some physical properties of these thin films appear to depend on the oxygen pressure during growth: the film grown at low oxygen pressure (P$\approx10^{-7}~mbar$) has a metallic aspect and is conducting with a positive Hall signal while those grown under higher oxygen pressures ($7\times10^{-5}\leq~P\leq0.4 mbar$) are insulating. However no intrinsic ferromagnetic signal could be attributed to the HfO$_2 $ films, irrespectively of the oxygen pressure during the deposition.
\end{abstract}


\submitto{\JPCM}

\maketitle

\section{Introduction}
The unexpected ferromagnetism at high temperature reported recently in several oxides of \textit{d} elements such as ZnO~\cite{ueda2001,venkatesan2004a}, TiO$_2$~\cite{matsumoto2001,shinde2003,griffin2005,hoahong2006}, HfO$_2$~\cite{venkatesan2004b,coey2005} or borides such as CaB$_2$C$_2$~\cite{akimitsu2001} and CaB$_6$~\cite{young1999,matsubayashi2002,young2002,lofland2003,dorneles2004,bennett2004} has triggered a considerable interest and posed an exciting fundamental problem. If a high Curie temperature in transition-metal doped wide-band-gap oxides such as ZnO can be predicted extrapolating the Zener model of carrier-induced ferromagnetism, the connection between magnetic properties (T$_C$) and doping has not yet been experimentally proved even in the widely studied 3\textit{d}-doped ZnO. The question about the origin of ferromagnetism in undoped CaB$_6$ or HfO$_2$ still remains open because the transition metal ions involved in these materials have neither open \textit{d} nor open \textit{f} electronic shells. To emphasize the difference between this unconventional ferromagnetism and the classical 3\textit{d} exchange ferromagnetism the term "\textit{d}$^0$ ferromagnetism" was introduced. The very first suggestion to explain the weak ferromagnetism in La doped CaB$_6$ single crystals was the freezing into a Wigner crystal of the low-density electron gas~\cite{ceperley1999}. Rapidly however, first-principles calculations have pointed out the role of point defects in the stabilization of high-temperature ferromagnetism in these systems~\cite{monnier2001,ichinomiya2002,elfimov2002,daspemmaraju2005}. However the exact nature of neither defects: extrinsic contamination~\cite{matsubayashi2002,lofland2003,bennett2004}, point defects~\cite{monnier2001}, intrinsic doping effect~\cite{ichinomiya2002}, anion or cation vacancies~\cite{elfimov2002,daspemmaraju2005}, nor exchange mechanism~\cite{ichinomiya2002,osorio2006,bouzerar2006} is yet completely settled. The magnetic moment carried by some defects in semiconductors is widely used in the electron paramagnetic resonance technique to study the nature and density of defects. This technique evidences volume and even surface magnetic states. However in these studies paramagnetic behaviour is recorded. In HfO$_2$, theoretical models based on vacancies-induced mechanisms predict that cation vacancies can exhibit a high-spin state with an associated magnetic moment as large as 4$~\mu_B$~\cite{daspemmaraju2005}. Recently Bouzerar and Ziman~\cite{bouzerar2006} have proposed a model that accounts more generally for the vacancy-induced \textit{d}$^0$ ferromagnetism in oxide compounds. In this model the cation vacancy or any substitutional defect creates an extended magnetic moment on neighboring oxygen atoms. They are able, by treating the randomness exactly, to calculate the magnetic couplings between these moments and thus the Curie temperature. They predict an enhanced ferromagnetic coupling between resonant impurity levels. This effect is strongly sensitive to the amount of impurities, thus small concentrations of defects can stabilize Curie temperatures well above room temperature.

Among all these oxides, HfO$_2$ is of particular interest because of potential applications for electronic devices. HfO$_2$ is a high-\textit{k} dielectric material ($\epsilon\approx25$) and a wide band gap oxide ($E_g\approx5.7$ eV~\cite{Balog1977}). Owing to its high thermal stability and low leakage current, it is a very promising candidate to replace silicon dioxide as a gate material for next generation silicon metal-oxide-semiconductor-field-effect transistors~\cite{ramanathan2004}. In an ionic picture, the Hf$^{4+}$ ion has a closed shell [Xe] 4\textit{f}$^{14}$ configuration and consequently it is nonmagnetic. Thus the observation of a ferromagnetic-like and strongly anisotropic behaviour, well above the room temperature, without magnetic 3\textit{d} element substitution such as Co or Mn (elements more often used in semiconductors) is very intriguing~\cite{venkatesan2004b,coey2005}. Actually if this property is confirmed, it would widen substantially its application in the field of spintronics. Furthermore, HfO$_2$ is a simple material and a good model to study the origin of this unconventional ferromagnetism.

Following the first claim of ferromagnetism in HfO$_2$~\cite{venkatesan2004b}, the same group reported a more detailed study on films of different thicknesses, obtained by pulsed-laser deposition under different conditions and on different substrates~\cite{coey2005}. For all these undoped films the Curie temperatures extrapolate far beyond 400 K. However it turns out that the value of the magnetic moment does not depend clearly on either the film thickness or the type of the substrate. Apparently the moments are unstable over extended periods of time, indeed a decrease of about 10\% of the moment is observed after 6 months. Later on several experimental studies on the magnetic properties of HfO$_2$ thin films have been reported in the literature with very controversial results. Hong \etal~\cite{hoahong2006} have confirmed the ferromagnetism above room temperature in 200 nm thick HfO$_2$ films also prepared by pulsed-laser deposition. On the other hand Abraham \etal~\cite{abraham2005} and Rao \etal~\cite{rao2006} did not observe ferromagnetism in undoped HfO$_2$ films. The former group~\cite{abraham2005} has shown that films of HfO$_2$ grown by metalorganic chemical vapor deposition (MOCVD), which thickness ranges from 5 to 40 $nm$, exhibit no evidence of ferromagnetism. It stressed however that contamination by handling them with stainless-steel tweezers results in a measurable magnetic signal and that the magnetic behavior is similar to the one reported in reference~\cite{coey2005} for HfO$_2$ films, including the magnitude of the moments, the magnetic field dependence and anisotropy. In the same way Rao \etal~\cite{rao2006} did not observe ferromagnetism in HfO$_2$ films grown by pulsed-laser deposition on (100) yttria stabilized ZrO$_2$ (YSZ) substrates. Though a ferromagnetic behavior is observed at room temperature in Co-doped HfO$_2$ films, its origin appears to be extrinsic, very likely due to the formation of a Co-rich surface layer. Finally in their model of vacancy-induced ferromagnetism, Bouzerar and Ziman~\cite{bouzerar2006} predict that, for usual densities of cation vacancies, HfO$_2$ is near the edge of stability of the ferromagnetism, but does not exhibit it. The stabilisation of ferromagnetism in HfO$_2$ could be realized by substituting an element of Group 1A on the Hf sites. The existence of ferromagnetism in hafnium oxide films, as well as its physical origin, remains still matter of debate.

In the present paper we report on detailed magnetic studies of HfO$_2$ thin films grown by pulsed laser deposition on \textit{c}-cut sapphire substrates under different partial oxygen pressures. The purpose of this work is to seek the eventual modifications of the magnetic behaviour of the HfO$_2$ films induced by the oxygen pressure. By changing the partial oxygen pressure from 0.4 to 1.2$\times10^{-7}~mbar$, we expect to increase the defect content due to oxygen but also hafnium vacancies in films having the same thickness.

\section{Experimental details and results}
\subsection{Film preparation}
The films were prepared by pulsed-laser deposition, using the tripled frequency ($\lambda=355~nm$) of a Nd:YAG laser; the fluence on the target is $10~J/cm^2$. The target is a $16~mm$-diameter HfO$_2$ pellet obtained by compressing a 99.995\%-pure Alfa-Aesar powder of hafnium oxide. According to the supplier analysis certificate, the different identified impurities are Zr ($<50~ppm$), Al ($<25~ppm$), Si ($<25~ppm$) and Ti ($<25~ppm$). Before installation into the deposition chamber, the pellet was annealed in air at $1400~^{\circ}C$ for 12 hours in a Pt crucible. The substrates are $5\times5\times0.5~mm^3$ \textit{c}-sapphire single crystals (masses of which vary between 47 and $63~mg$). During the deposition their temperature was fixed at $750~^{\circ}C$. Four films were prepared under oxygen pressures ranging from $1.2\times10^{-7}$ to $4\times10^{-1}~mbar$ (see Table~\ref{table1}). The film thickness, estimated in situ by monitoring oscillations of the reflectivity during deposition and taking into account the 2.0 optical index of HfO$_2$ at $670~nm$, is of about $50~nm$. All the films have a transparent colorless aspect except film 1, which was deposited under the lowest oxygen pressure, $1.2\times10^{-7}~mbar$ (see Table~\ref{table1}). This film presents a dark grey metallic aspect, characteristic of a strong reduction of HfO$_2$.

\subsection{Structural characterization}
The crystalline structure of the films was investigated by x-ray diffraction using a Siemens D5000 diffractometer with Co $K_\alpha$ radiation. The diffraction diagram of the initial powder, checked using the same diffractometer, is fully consistent with the HfO$_2$ monoclinic structure, $P~2_1/c$ space group, and values of the lattice parameters: \textit{a} = 5.1156~\AA, \textit{b} = 5.1722~\AA, \textit{c} = 5.2948~\AA, $\beta = 99.18^{\circ}$, reported in the ICSD tables~\cite{ICSD}. Due to the strong texture of the films only few reflections are fairly well resolved in the diffraction patterns as illustrated in figure~\ref{figure1}. Moreover the more the oxygen pressure is low the more the crystalline quality of the film is poor. Cell parameters could be refined for film 4 (P = $4\times10^{-1}~mbar$), which has the best crystalline quality and gives more reflections. Refined values are: \textit{a} = 5.116(3)~\AA, \textit{b} = 5.151(3)~\AA, \textit{c} = 5.285(3)~\AA, $\beta$ = $98.9(2)^{\circ}$, in good agreement with the HfO$_2$ monoclinic structure reported above. Note that the volume of the HfO$_2$ films corresponds roughly to 0.01\% of the whole sample volume (substrate + film). No other peaks are detectable in the x-ray diffraction patterns. This let suppose that, if they are present, parasitic phases are not textured or amorphous.

\subsection{Magnetic measurements}
The magnetic characterization of the films was carried out on a MPMS Quantum Design SQUID magnetometer using the reciprocating sample option (RSO). The magnetic field is applied parallel to the plane of the films and the signal recorded with the sample oscillating around the center of the SQUID pickup-coils. The signal is fitted to an ideal dipole response using a non-linear least-squares routine. In this configuration the magnetometer manufacturer gives an absolute sensitivity of 1$\times10^{-11} Am^2$ (1$\times10^{-8} emu$) at $0.25~T$ and an accuracy of 0.1\% for a standard sample (cylinder of $\phi=3~mm,~h=3~mm$). The actual accuracy is less good. Indeed it is limited by \textit{i}) errors relative to the sample shape and dimensions, which differ from those of the standard, \textit{ii}) errors coming from the sample mounting: radial off-centering and/or angular misalignment, \textit{iii}) other instrumental errors due to non reproducible factors not controlled by the experimentalist. For present measurements we estimate an accuracy of the calibration factor of 5\% (i.e. an absolute error of $1\times10^{-7}~Am^2$ for a signal of $2\times10^{-6}~Am^2$).

The back of the substrates has been scrubbed with sandpaper in order to remove the silver paint, used for thermal contact during deposition. The sample was then rinsed in acetone before being mounted into a sample holder made out of a plastic straw. As the dimensions of the samples are smaller than the straw diameter, they are wrapped in a thin plastic membrane that allows holding in position the sample inside the straw. It was checked, by measuring the empty plastic membrane mounted inside the straw with exactly the same geometry as when measuring the samples, that it gives no detectable signal. The four samples have been characterized in a systematic way performing the same sequence of measurements. Magnetization processes, $M(H)$, in fields up to $5~T$ were measured at 2, 5, 10 and 300 $K$. The thermal variation of the magnetic moment, $M(T)$ was measured in the temperature range $2-300 K$ under an applied field of $0.05~T$. Finally magnetization loops were performed between $\pm2~T$ at 2 and 300 $K$. After these magnetic measurements have been fully completed, the HfO$_2$ film on sample 1 was totally removed by argon ion etching (stripped sample in the following). Then, exactly the same sequence of measurements as that used previously for the four samples, was performed on the stripped sample. With these measurements it is expected to determine, as accurately as possible, the contribution of the substrate and silver paint scraps to the magnetic signal.

$M(H)$ measurements performed on the Alfa-Aesar powder have confirmed a diamagnetic behaviour of the HfO$_2$ powder. The value of the HfO$_2$ diamagnetic susceptibility deduced from these measurements is $\chi=-1.07\times10^{-9}~m^3/kg$ (an overestimated value of $\chi=-2.8\times10^{-9}~m^3/kg$ can be deduced from the theoretical values of the Hf$^{4+}$ and O$^{2-}$ ion diamagnetic susceptibilities~\cite{Landolt-Börnstein1979}). Assuming that the HfO$_2$ film has the same density than bulk HfO$_2$, a film of dimensions $5~mm\times5~mm\times50~nm$ will have a mass of $1.27\times10^{-2}~mg$ and then a magnetic signal of $6.8\times10^{-17}~Am^2$ under a field of $5~T$. This value is far below the instrumental sensitivity.

Within the experimental accuracy, the $M(H)$ curves for the four samples and the stripped one, do not show any temperature dependence, we thus report in figure~\ref{figure2} the magnetization processes measured at 2 $K$, the lowest temperature. It is obvious that the five curves are characteristic of a diamagnetic behaviour of all the samples with no evidence of ferromagnetic signal. The experimental data compare very well with the signal calculated for a sapphire sample having the same mass as the considered sample. The sapphire signal was calculated using the value of the diamagnetic susceptibility, $\chi_{sapphire}=-4.4\times10^{-9}~m^3/kg$, obtained from measurements on blank substrates~\cite{fillion2007}. This value is not very different from the one given in reference~\cite{coey2005}, $\chi_{sapphire}=-4.6\times10^{-9} m^3/kg$. The value of the magnetic signal and its field variation are exactly the same for the four samples and the stripped one. It is worth noting that the five curves compares pretty well with the one of the blank sapphire substrate reported by Coey~\etal~\cite{coey2005}.

The magnetization loops between -0.2 and 0.2~$T$ reveal the opening of an extremely small hysteresis cycle at all temperatures and for all the samples even for the stripped sample. Hysteresis loops may arise from instrumental hysteresis, due for instance to remanent fields in the superconducting coil. In this low field range, they can also be artifacts due to the mathematical fit by the least-squares routine of very weak signals. That the hysteresis loops for all samples are more open at low temperatures could let suppose the existence of a weak contribution of "ferromagnetic"-type in addition to the main diamagnetic signal of the samples. The fact that the hysteresis loop at 2 $K$ is larger for the stripped sample than for the four HfO$_2$ film-deposited samples (see for instance the comparison between sample 1 and the stripped sample in figure~\ref{figure3}) strongly supports that the hysteresis is not associated to the HfO$_2$ films. An estimation of this contribution can be performed by subtracting the main diamagnetic contribution deduced from the magnetization processes under high fields (see figure~\ref{figure2}). The remaining signal is twice larger in the stripped sample than in the four HfO$_2$ film-deposited samples. This clearly evidences that the "ferromagnetic"-type contribution cannot be ascribed to the HfO$_2$ films. Despite the hysteresis loop, which can be due to experimental artifacts, as said above, this remaining signal is more reminiscent of a paramagnetic-type signal than of a ferromagnetic-type one, as illustrated by panels c and d in figure~\ref{figure3}. It is very likely that this contribution, characterized by a positive susceptibility, arises from impurities that can contaminate the substrate itself. The contamination can also result from sample handling. The stripped sample is probably the more polluted one, because it has been more handled than the other samples.

Figure~\ref{figure4} shows the thermal dependence of the magnetic moment, $M(T)$, under an applied field of 0.05~$T$ for the four HfO$_2$ film-deposited samples and the stripped sample. A negative magnetic moment is observed, almost constant from $300~K$ down to about 50$~K$. On further lowering of temperature the signal shows a clear upturn and starts increasing. This thermal variation is consistent with the superposition of a weak "paramagnetic-type" contribution to the main temperature-independent diamagnetic one. Whatever the origin of this "paramagnetic" signal may be, assuming it follows a Curie law with an effective Curie constant for each sample, the product $M(T)\times{T}=C_{eff}\times{H}+M_{dia}\times{T}$, should vary linearly with temperature and the value of the diamagnetic moment is given by the slope. As shown in figure~\ref{figure5}, a linear variation is effectively observed in all the samples, the four HfO$_2$ film-deposited and the stripped sample. The values of the diamagnetic contribution under $0.05~T$ deduced from the linear fit of $M(T)\times{T}$ curves are reported in table~\ref{table1}. They compare very well with the signals calculated for sapphire samples of same mass, with $\chi_{sapphire}=-4.4\times10^{-9}~m^3/kg$, except for sample 4. For this sample the difference between the experimental and calculated values is larger than the experimental error.

In very weak magnetic or non-magnetic systems, paramagnetic tails are very often observed at low temperatures in $M(T)$ measurements. These tails are due to small amounts of diluted magnetic impurities that remain paramagnetic. This pollution can be hardly avoided even in very controlled atmospheres and using highly purified elements. At this stage the nature of the impurities in the five samples studied here remains unknown. But again the "paramagnetic-type" contribution is not related to magnetic moments arising from the HfO$_2$ films, since the HfO$_2$ film-deposited samples and the stripped sample present the same tail (see figures~\ref{figure4} and~\ref{figure5}). As iron and iron oxides are actually the most frequent source of magnetic impurities, it was tried to account for the low temperature tail assuming Fe$^{2+}$ impurities. For samples 1, 2, 3 and for the stripped sample, as shown in figure~\ref{figure4}, the evolution of the moment at low temperatures is well reproduced by a content of about $10^{15}$ Fe$^{2+}$ ion impurities ($\mu_{eff}(Fe^{2+})=5.4~\mu_B$). Note that the paramagnetic signal calculated using this amount of Fe$^{2+}$ impurities between -0.2 and 0.2~$T$ is consistent with the remaining signal in the magnetization loops. This is illustrated, for example in figure~\ref{figure3} for the sample 1 and the stripped sample. It is worth noting that the result would be not very different for Fe$^{3+}$ ions ($\mu_{eff}(Fe^{3+})=5.9~\mu_B$). $10^{15}$ Fe ions would represent a mass of about $9\times10^{-8}~g$ and a volume concentration of $\approx1~ppm$ of the substrate volume and $\approx1\%$ of the HfO$_2$ film volumes. Suppose that the $10^{15}$ Fe ions form a perfectly crystallized film its thickness would correspond to about 4 atomic planes ($\approx0.5~nm$), this quantity can be hardly observed with our experimental x-ray diffraction (XRD) setup. Fits have been performed on the supposition that the "paramagnetic-type" signal comes from Fe impurities, but pollution by any other magnetic ions is also likely. Whatever the impurities, the fact that the stripped sample can be fitted with the same amount of impurities than the HfO$_2$ film-deposited samples strongly suppose that they are scattered inside the substrate (trivalent transition-metal cations are very often present inside sapphire).

For sample 4 the experimental $M(T)$ curve is not reproduced as well as for the other samples as shown in figure~\ref{figure6}. Since the diamagnetic contribution deduced from the fit of $M(T)\times{T}$ is smaller than the one expected for the substrate (table~\ref{table1}), it can be supposed that this difference comes from a positive "ferromagnetic"-type contribution. This contribution, estimated at $\approx2.3\times10^{-9}~Am^2$ at 2~$K$, is three times smaller than the diamagnetic contribution under $0.05~T$ (see table~\ref{table1}). If this moment is linked with "\textit{d}$^0$ ferromagnetism" in the HfO$_2$ film, its value is more than one order in magnitude smaller than those reported in reference~\cite{coey2005} for films of same dimensions and grown under very similar conditions. This moment is even smaller than the moment ($5\times10^{-9}~Am^2$) measured at room temperature on a sample after removing the HfO$_2$ film~\cite{coey2005}. In the same work~\cite{coey2005} it was reported that HfO$_2$ powder heated in vacuum develops a weak magnetic moment. This moment that was attributed to the magnetic moment induced by the oxygen vacancies is eliminated on annealing in oxygen. The HfO$_2$ film of sample 4 has been deposited under a high oxygen pressure ($0.4~mbar$), thus the apparition of a ferromagnetic signal in the film would be in contradiction with these previous observations. An other conceivable possibility is the formation of the ferrimagnetic $\gamma$-Fe$_2$O$_3$ phase during the deposition. In this case a moment of $2.3\times10^{-9}~Am^2$ could correspond to $\approx3\times10^{-8}~g$ ($\approx1\times10^{14}$ formulas) of Fe$_2$O$_3$. Taking the density of bulk $\gamma$-Fe$_2$O$_3$ (5.07$~g/cm^3$) this leads to a volume of $\approx6\times10^{-15}~m^3$ (volume concentration of $\approx0.5~ppm$ of the substrate volume). Here also this quantity would be hardly detected by XRD unless the phase is very well crystallized and textured, which is unlikely. Though we cannot predict the shape, size and distribution of the $\gamma$-Fe$_2$O$_3$ precipitates within the sample, the total volume is large enough to allow precipitates with ferromagnetic behaviour, indeed the diameter limit for the onset of superparamagnetism in spheroidal $\gamma$-Fe$_2$O$_3$ particles is around 6-7~$nm$~\cite{coey1972}.

\subsection{Magneto-transport measurements}
The HfO$_2$ film of sample 1 prepared under an oxygen pressure of $10^{-7}~mbar$ presents a metallic aspect. In order to characterize its transport properties Hall measurements were performed between $300~K$ and $3~K$ under applied magnetic fields varying between -6 to $6~T$. The Hall voltage was measured in the Van der Pauw configuration in dc-current mode with a current of $0.1~mA$, directly contacting the sample with mechanical probes. In the whole temperature range the Hall voltage presents a linear variation with magnetic field and no evidence of any anomalous Hall effect as illustrated in figure~\ref{figure7}, and in agreement with magnetic measurements. The sign of the Hall voltage with respect to the signs of current and magnetic field is characteristic of \textit{p}-type carriers. Assuming a thickness of $50~nm$ for the film it was estimated that the carrier density is of the order of $10^{22}~holes/cm^3$. This carrier density is two orders of magnitude larger than in magnetic semiconductors such as (Ga,Mn)As or \textit{p}-(Zn,Mn)Te~\cite{dietl2000}. Despite this high density, a rough estimate of the carrier mobility gives $10^{-4}~m^2/V.s$. Such a poor mobility, two hundred times smaller than in epitaxial ZnTe with usual doping, could be ascribed to a high density of defects in the film structure as supported by the poor x-ray diffraction quality. A theoretical study of vacancy and interstitial defects in hafnia predicts that oxygen vacancies introduce electronic levels in the band gap of HfO$_2$~\cite{foster2002}. These levels are situated around $2.8~eV$ above the top of the valence band. This study is limited to a small amount of oxygen vacancies, which is certainly not the case in the strongly reduced HfO$_2$ film. However according to this results, the observation of \textit{p}-type carriers in HfO$_2$ film of sample 1 would indicate that a high rate of oxygen vacancies pulls down the levels closer to the top of the valence band. This could hint at the appearance of a conducting impurity band in the gap with hole character. It was checked that the \textit{p}-type conductivity is a reproducible property of HfO$_2$ films prepared under very low oxygen pressure, P $\approx~1\times10^{-7}~mbar$.

\section{Conclusion}
We have prepared thin ($\approx50~nm$ thick) films of HfO$_2$ by pulsed-laser deposition on \textit{c}-sapphire crystal under four different partial pressures of oxygen between $1.2\times10^{-7}$ and $4\times10^{-1}~mbar$. Many studies support that cation vacancies may induce ferromagnetism in oxides. Though in such films it is very difficult to properly characterize the defects, it is likely that oxygen vacancies induced by tuning the partial pressure during the deposition of the HfO$_2$ films also influences the density of hafnium vacancies. Present work shows that none of the four studied HfO$_2$ films presents ferromagnetic properties that could be linked to "\textit{d}$^0$ ferromagnetism". Under low fields and at low temperatures, the different features that let suppose magnetic contributions other than the diamagnetic signal from the substrate are well accounted for by assuming the existence of very weak contaminations by magnetic ions such as iron ions and/or $\gamma$-Fe$_2$O$_3$ precipitates. The upturn of the magnetic moment at the lowest temperatures, observed in the $M(T)$ curves for all the samples even for the stripped one, is characteristic of a contamination by paramagnetic impurities, i.e. diluted pollution of isolated atoms (ions) or clusters smaller than few $nm$. Though the nature of these impurities in the five samples remains unknown, this tail is well reproduced by an assembly of about $10^{15}$ non interacting Fe$^{2+}$ ions. For sample 4, prepared under the highest partial oxygen pressure (0.4 $mbar$), the slight excess of positive signal is in all likelihood due to the presence of a small amount of a ferromagnetic such as the $\gamma$-Fe$_2$O$_3$ phase. Sources of pollution of the samples are numerous. Even in very controlled atmospheres impurities are always present in the deposition chamber. Impurities remain present inside the substrates themselves inherent in the manufacture. Finally handling the samples leads also to pollution. This points out the difficulties encountered when dealing with tiny ferromagnetic signals, which turn to be of same magnitude than those coming from any impurity or defect in the sample. Our results are consistent with previous experimental works reported by Rao~\etal~\cite{rao2006} and Abraham~\etal~\cite{abraham2005}. They are also in agreement with the calculations of Bouzerar and Ziman~\cite{bouzerar2006} that predict no ferromagnetism in pure HfO$_2$. The most unexpected result is that a strong reduction of a $50~nm$-thick HfO$_2$ film induces \textit{p}-type carriers. Studies are underway in order to understand the origin of this effect.

\subsection{Acknowledgments}
The authors would like to acknowledge our colleague P. Bordet and X. Chaud for their kind assistance in the course of the present study. This work was supported by the cooperation program between CNRS and Russian Academy of Sciences, and RFBR, grant No.04-02-16191.

\newpage
\begin{table}
\begin{center}
\caption{\label{table1} Oxygen pressure during the deposition of the films, total mass (film and substrate), and appearance of the samples. The diamagnetic contributions of the samples deduced from the fit of $M(T)\times{T}$ curves and calculated for pure sapphire samples of the same mass with $\chi_{sapphire}=-4.4\times10^{-9} m^3/kg$ are reported in rows 5 and 6 respectively.}

\begin{tabular}{cccccc}
\br
Sample & $O_2$~pressure & mass & appearance & $M_{dia}(0.05~T)$ & $M_{sapphire}(0.05~T)$ \\ 
    &($mbar$)&($mg$)&    &($Am^2$)&($Am^2$) \\
\br
 1 & $1.2\times10^{-7}$ & 50$\pm1$ & grey metallic & $-9.23\times10^{-9}$ & $-8.75\times10^{-9}$ \\ 
\mr
 2 & $7\times10^{-5}$ & 63$\pm1$ & transparent colorless & $-11.00\times10^{-9}$ & $-10.98\times10^{-9}$ \\ 
\mr
 3 & $1.2\times10^{-2}$ & 52$\pm1$ & transparent colorless & $-9.39\times10^{-9}$ & $-8.98\times10^{-9}$ \\ 
\mr
 4 & $4\times10^{-1}$ & 47$\pm1$ & transparent colorless & $-6.55\times10^{-9}$ & $-8.19\times10^{-9}$ \\ 
\mr
 Stripped sample & - & 50$\pm1$ & transparent & $-8.50\times10^{-9}$ & $-8.75\times10^{-9}$ \\ 
\br
\end{tabular}
\end{center}
\end{table}

\Figures
\Figure{\label{figure1} Diffraction diagram of sample 4 (Co $K_\alpha$ radiation).}

\Figure{\label{figure2} Magnetization processes measured at 2~$K$ of samples 1, 2, 3, 4 and the stripped sample. The full lines represent the diamagnetic contribution of a sapphire sample having the same mass as the measured sample, calculated using the value of the sapphire susceptibility, $\chi_{sapphire}=-4.4\times10^{-9}~m^3/kg$~\cite{fillion2007}.}

\Figure{\label{figure3} Upper panels, raw data of magnetization cycles at 2 $K$ for (a) sample 1, (b) the stripped sample. Bottom panels show the difference between the experimental data and the main diamagnetic contribution deduced from the magnetization processes under high fields, for sample 1 (c) and the stripped sample (d) respectively. Full lines show the paramagnetic signal calculated using an amount of $2.1\times10^{15}$ (c) and $3.1\times10^{15}$ (d) Fe$^{2+}$ impurities.}

\Figure{\label{figure4} Thermal variation of the magnetic moment of samples 1, 2, 3, 4 and the stripped sample, measured under a field of 0.05 $T$, applied parallel to the plane. Open symbols represent the experimental data. Full lines show the calculated thermal variation assuming \textit{i}) a temperature independent diamagnetic contribution, \textit{ii}) a paramagnetic contribution of N Fe$^{2+}$ cations.}

\Figure{\label{figure5} $M(T)\times{T}$ as a function of the temperature deduced from experimental curves under 0.05 $T$(open dots) and linear fits (lines) for the four samples and the stripped one.}

\Figure{\label{figure6} Thermal variation of the moment for sample 4 ($\mu_0H=0.05~T$). The open dots represent the experimental data. The dashed line represents the sum of the diamagnetic contribution determined from the fit of the $M(T)\times{T}$ curve and the calculated paramagnetic contribution of $6.2\times10^{14}$ Fe$^{2+}$ ions. The difference between the experimental curve and the calculations (full line) would correspond to a weak  contribution ($\approx2.3\times10^{-9}~Am^2$) of "ferromagnetic"-type.}

\Figure{\label{figure7} Sample 1: Hall resistivity $\rho_{xy}$ of the HfO$_2$ film as a function of the applied magnetic field obtained in the Van der Pauw configuration with a current of 0.1~$mA$ and for a 50~$nm$-thick film. Open circles (respectively solid squares) are the experimental values at 300~$K$ (3~$K$). The dot and solid lines represent the corresponding linear fits.}


\section{References}
\begin{thebibliography}{99}
\bibitem{ueda2001} Ueda K, Tabat H and Kawai T 2001 \textit{Appl. Phys. Lett.} \textbf{79} 988.   
\bibitem{venkatesan2004a} Venkatesan M, Fitzgerald C B, Lunney J G and Coey J M D 2004 \textit{Phys. Rev. Lett.} \textbf{93} 177206.
\bibitem{matsumoto2001} Matsumoto Y, Murakami M, Shono T, Hasegawa T, Fukumura T, Kawasaki M, Ahmet P, Chikyow T, Koshihara S and Koinuma H 2001 \textit{Science} \textbf{291} 1056.
\bibitem{shinde2003} Shinde S R, Ogale S D, Das Sarma S, Simpson J R, Drew H D, Lofland S E, Lanci C, Buban J P, Browning N D, Kulkarni V N, Higgins J, Sharma R P, Greene R L and Venkatesan T 2003 \textit{Phys. Rev. B} \textbf{67} 115211.
\bibitem{griffin2005} Griffin K A, Pakhomov A B, Wang C M, Heald S M and Krishnan K M 2005 \textit{Phys. Rev. Lett.} \textbf{94} 157204.
\bibitem{hoahong2006} Hoa Hong N, Sakai J, Poirot N and Briz\'{e} V 2006 \textit{Phys. Rev. B} \textbf{73} 132404.
\bibitem{venkatesan2004b} Venkatesan M, Fitzgerald C B and Coey J M D 2004 \textit{Nature (London)} \textbf{430} 630.
\bibitem{coey2005} Coey J M D, Venkatesan M, Stamenov P, Fitzgerald C B and Dorneles L S 2005 \textit{Phys.Rev.B} \textbf{72} 024450.
\bibitem{akimitsu2001} Akimitsu J, Takenawa K, Suzuki K, Harima H and Kuramoto Y 2001 \textit{Science} \textbf{293} 1125.
\bibitem{young1999} Young D P, Hall D, Torelli M E, Fisk Z, Sarrao J L, Thompson J D, Ott H R, Oseroff S B, Goodrichk R G and Zysler R 1999 \textit{Nature (London)} \textbf{397} 412.
\bibitem{matsubayashi2002} Matsubayashi K, Maki M, Tsuzuki T, Nishioka T and Sato N K 2002 \textit{Nature} \textbf{420} 143.
\bibitem{young2002} Young D P, Fisk Z, Thompson J D, Ott H R, Oseroff S B and Goodrich R G 2002 \textit{Nature} \textbf{420} 144.
\bibitem{lofland2003} Lofland S E, Seaman B, Ramanujachary K V, Hur N and Cheong S W 2003 \textit{Phys. Rev. B} \textbf{67} 20410.
\bibitem{dorneles2004} Dorneles L S, Venkatesan M, Moliner M, Lunney J G and Coey J M D 2004 \textit{Appl. Phys. Lett.} \textbf{85} 6377.
\bibitem{bennett2004} Bennett M C, van Lierop J, Berkeley E M, Mansfield J F, Henderson C, Aronson M C, Young D P, Bianchi A, Fisk Z, Balakirev F and Lacerda A 2004 \textit{Phys. Rev. B} \textbf{69} 132407.
\bibitem{ceperley1999} Ceperley D 1999 \textit{Nature (London)} \textbf{397,} 386.
\bibitem{monnier2001} Monnier R and Delley B 2001 \textit{Phys. Rev. Lett.} \textbf{87} 157204.
\bibitem{ichinomiya2002} Ichinomiya T 2002 \textit{Phys. Rev. B} \textbf{63} 045113.
\bibitem{elfimov2002} Elfimov I S, Yunoki S and Sawatzky G A 2002 \textit{Phys. Rev. Lett.} \textbf{89} 216403.
\bibitem{daspemmaraju2005} Das Pemmaraju C and Sanvito S 2005 \textit{Phys. Rev. Lett.} \textbf{94} 217205.
\bibitem{osorio2006} Osorio-Guillen J, Lany S, Barabash S V and Zunger A 2006 \textit{Phys. Rev. Lett.} \textbf{96} 107203.
\bibitem{bouzerar2006} Bouzerar G and Ziman T 2006 \textit{Phys. Rev. Lett.} \textbf{96} 207602.
\bibitem{Balog1977} Balog M, Schieber M, Michiman M and Patai S 1977 \textit{Thin Solid Films} \textbf{41} 247.
\bibitem{ramanathan2004} Ramanathan S, Mclyntyre P C, Guha S and Gusev E 2004 \textit{Appl.Phys.Lett.} \textbf{84} 389.
\bibitem{abraham2005} Abraham D W, Frank M M and Guha S 2005 \textit{Appl. Phys. Lett.} \textbf{87} 252502.
\bibitem{rao2006} Rao M S R, Kundaliya D C, Ogale S B, Fu L F, Welz S J, Browning B D, Zaitsev V, Varughese B, Cardoso C A, Curtin A, Dhar S, Sginde S R, Venkatesan T, Lofland S E and Schwarz S A 2006 \textit{Appl. Phys. Lett.} \textbf{88}, 142505.
\bibitem{ICSD} ICSD card 27313 (1986).
\bibitem{fillion2007}Fillion G, \textit{private communication}
\bibitem{Landolt-Börnstein1979} Landolt-Börnstein  1979 \textit{Magnetic propertie of transition metal compounds}, vol~II/10 supplement~2, Hellwege K H and Hellwege A M Editors, Springer-Verlag.
\bibitem{coey1972} Coey J M D and Khalafalla D 1972 \textit{Phys. Stat. Sol. a} \textbf{11} 229.
\bibitem{dietl2000} Dietl T, Ohno H, Matsukura F, Cibert J and Ferrand D 2000 \textit{Science} \textbf{287} 1019.
\bibitem{foster2002} Foster A S, Lopez Gejo F, Shluger A L and Nieminen R M 2002 \textit{Phys. Rev. B} \textbf{65} 174117.
\end{thebibliography}
\end{document}